\documentstyle[12pt,psfig]{article}
\newcommand{\beq}{\begin{equation}}
\newcommand{\eeq}{\end{equation}}
\newcommand{\bea}{\begin{eqnarray}}
\newcommand{\eea}{\end{eqnarray}}

\renewcommand{\d}{\delta}

\renewcommand{\b}{\beta}

\newcommand{\m}{\mu}

\newcommand{\s}{\sigma}
\newcommand{\D}{\Delta}
\newcommand{\th}{\theta}

\newcommand{\oh}{\frac{1}{2}}

\newcommand{\non}{\nonumber}

\newcommand{\rf}[1]{(\ref{#1})}
\newcommand{\ra}{\rightarrow}

\newcommand{\pa}{\partial}

\begin{document}

\addtolength{\baselineskip}{0.20\baselineskip}

\hfill hep-lat/9610005

\hfill LBNL-39424

\hfill SWAT-96-119

\hfill September 1996

\begin{center}

\vspace{32pt}

{ {\large \bf Center Dominance and $Z_2$ Vortices In \\ $SU(2)$ 
              Lattice Gauge Theory } }

\end{center}

\vspace{18pt}

\begin{center}
{\sl L. Del Debbio${}^a$, M. Faber${}^b$, J. Greensite${}^{c,d}$,
and {\v S}. Olejn\'{\i}k${}^e$}

\end{center}

\vspace{18pt}

\begin{tabbing}

{}~~~~~~~~~~~~~\= blah  \kill
\> ${}^a$ Department of Physics, University of Wales Swansea, \\
\> ~~SA2 8PP Swansea, UK.  E-mail: {\tt L.Del-Debbio@swansea.ac.uk} \\
\\
\> ${}^b$ Institut f\"ur Kernphysik, Technische Universit\"at Wien, \\
\> ~~1040 Vienna, Austria.  E-mail: {\tt faber@kph.tuwien.ac.at} \\
\\
\> ${}^c$ Physics and Astronomy Dept., San Francisco State University, \\
\> ~~San Francisco CA 94132 USA.  E-mail: {\tt greensit@stars.sfsu.edu} \\
\\
\> ${}^d$ Theoretical Physics Group, Ernest Orlando Lawrence Berkeley \\
\> ~~National Laboratory, University of California, Berkeley, \\
\> ~~California 94720 USA.  E-mail: {\tt greensite@theorm.lbl.gov} \\
\\
\> ${}^e$ Institute of Physics, Slovak Academy of Sciences, \\
\> ~~842 28 Bratislava, Slovakia.  E-mail: {\tt fyziolej@savba.savba.sk} \\

\end{tabbing}

\newpage

\noindent {~}

\vspace{18pt}

\begin{center}

{\bf Abstract}

\end{center}

\bigskip

   We find, in close analogy to abelian dominance in maximal abelian
gauge, the phenomenon of center dominance in maximal center gauge for 
$SU(2)$ lattice gauge theory.  Maximal center gauge is a gauge-fixing 
condition that preserves a residual $Z_2$ gauge
symmetry; ``center projection'' is the projection of $SU(2)$ link variables
onto $Z_2$ center elements, and ``center dominance'' is the fact that
the center-projected link elements carry most of the information about 
the string tension of the full theory.  We present numerical evidence that
the thin $Z_2$ vortices of the projected configurations are associated
with ``thick'' $Z_2$ vortices in the unprojected configurations.
The evidence also suggests that the thick $Z_2$ vortices
may play a significant role in the confinement process.

\vfill

\newpage

\section{Introduction}

   Perhaps the most popular theory of quark confinement is the
dual-superconductor picture as formulated in abelian-projection gauges 
\cite{tH2}. In this theory the unbroken $U(1)^{N-1}$ symmetry
of the full $SU(N)$ gauge group plays a special role
in identifying both the relevant magnetic monopole configurations, and  
also the abelian charge which is subject to the confining force.
The phenomenon of abelian dominance \cite{Suzuki} in maximal abelian
gauge \cite{KLSW} is often cited as strong evidence in favor
of the dual-superconductor picture (c.f. ref. \cite{Poli} for a
recent review).

   Of course, many alternative explanations of quark confinement have
been advanced over the years.  The theory which will concern us in
this article is the vortex condensation (or ``spaghetti vacuum'') picture, 
in which the vacuum is understood to be a condensate of vortices of some 
finite thickness, carrying flux in the center of the gauge group.  The 
spaghetti picture was originally advanced by
Nielsen and Olesen \cite{Poul}, and this idea was further elaborated 
by the Copenhagen group in the late seventies.  A closely related idea,
due to 't Hooft \cite{tH1} and Mack \cite{Mack}, emphasized
the importance of the $Z_N$ center of the $SU(N)$ gauge group.  
In that picture there is a certain correspondence 
between magnetic flux of the relevant vortices and the elements of the
the $Z_N$ subgroup, and it is random fluctuations in the number of such 
vortices linked to a Wilson loop which explains the area-law 
falloff.\footnote{See also ref. \cite{Yoneya}.  Recent work along these 
same lines is found in ref. \cite{Tomboulis}.}
Ref. \cite{AO} presents an argument for this $Z_N$ center restriction
in the framework of the ``Copenhagen vacuum.'' 

   The vortex condensation theory, like dual-superconductivity, focuses on
on a certain subgroup of the full $SU(N)$ gauge group, but 
it is the $Z_N$ center, rather than $U(1)^{N-1}$, which is 
considered to be of special importance.  This raises a natural question:
Does there exist, in close analogy to abelian dominance, some version of
``center dominance?''  If so, should this evidence be interpreted essentially
as a critique of abelian dominance, or should it be viewed as genuine
support for the vortex theory?   Supposing that the vortex condensation
theory is taken seriously, how can one identify $Z_2$ vortices
in unprojected field configurations, and can one determine 
if such vortices are of any physical importance?  This article is 
intended as a preliminary investigation of these questions.

\section{Center Dominance}

   We begin with the phenomenon of ``center dominance'' in maximal
center gauge.  One starts by fixing to the maximal abelian gauge 
\cite{KLSW}, which, for $SU(2)$ gauge theory, maximizes the quantity
\beq
      \sum_x \sum_{\m=1}^4 \mbox{Tr}[\s_3 U_\m(x) \s_3 U^\dagger_\m(x)]
\eeq
This gauge has the effect of making link variables as diagonal as possible,
leaving a remnant $U(1)$ gauge symmetry.  ``Abelian projection'' means the
replacement of the full link variables $U$ by the abelian links $A$,
according to the rule
\beq
      U = a_0 I + i\vec{a} \cdot \vec{\s} ~~~ \longrightarrow ~~~
      A = {a_0 I + i a_3 \s^3 \over \sqrt{a_0^2 + a_3^2} }
\label{proj}
\eeq
It can be shown that the $A$ link variables transform like $U(1)$ gauge
fields under the remnant $U(1)$ symmetry.  Abelian dominance, found by
Suzuki and collaborators \cite{Suzuki}, is essentially the fact that
the confining string tension can be extracted from the abelian-projected
$A$-link variables alone.  Abelian dominance has been widely interpreted
as supporting the dual-superconductor theory advanced
in ref. \cite{tH2}.  

   But while the dual-superconductor idea focuses on the remnant $U(1)$
subgroup of the gauge symmetry, it is the $Z_2$ center of the $SU(2)$ gauge
group that seems most relevant in the vortex condensation picture.
This suggests making a further gauge-fixing, which would bring the abelian
links as close as possible to the center elements $\pm I$ of $SU(2)$.
Therefore, writing
\beq
           A = \left[ \begin{array}{cc}
                      e^{i\th} &  \\
                               & e^{-i\th} \end{array} \right]
\label{A}
\eeq
we use the remnant $U(1)$ symmetry to maximize
\begin{equation}
      \sum_x \sum_\m \cos^2(\th_\m(x))
\label{maxZ2}
\end{equation}
leaving a remnant $Z_2$ symmetry.  This we call ``Maximal Center Gauge.''
Then define, at each link,
\begin{equation}
       Z \equiv \mbox{sign}(\cos \th) = \pm 1
\label{cp}
\end{equation}
which transforms like a $Z_2$ gauge field under the remnant symmetry.
``Center Projection'' $U \ra Z$, analogous to ``abelian projection''
$U \ra A$, is defined as the replacement of the full link variables $U$ 
by the center element $ZI$, in the computation of observables such as 
Wilson loops and Polyakov lines.

    Figure 1 is a plot of Creutz ratios vs. coupling $\b$, extracted from
Wilson loops formed from the center-projected $Z_2$ link variables.
Lattice sizes were $10^4$ for $\b \le 2.3$, $12^4$ at $\b=2.4$, and 
$16^4$ at $\b=2.5$.\footnote{Finite size effects, as indicated by the values
of center-projected Polyakov lines, appear to be significantly larger for
center-projected configurations as compared to the full link variables,
and this is why we use a $16^4$ lattice at $\beta=2.5$.}  What is rather
striking about Fig. 1 is the fact that, from 2 lattice spacings onwards,
the Creutz ratios at fixed $\b \ge 2.1$ all fall on top of one another,
and all lie on the same scaling line  
\beq
     \s a^2 = {\s \over \Lambda^2}({6\over 11}\pi^2 \b)^{102/121}
                 \exp[-{6\over 11}\pi^2 \b]
\eeq
with the value $\sqrt{\s}/\Lambda = 67$.  Even the logarithm of the
one-plaquette loop, $\chi(1,1)$, appears to parallel this line.  This
behavior is in sharp contrast to Creutz ratios extracted from the full
link variables, where only the envelope of Creutz ratios fits the
scaling line.

\begin{figure}
\centerline{\hbox{\psfig{figure=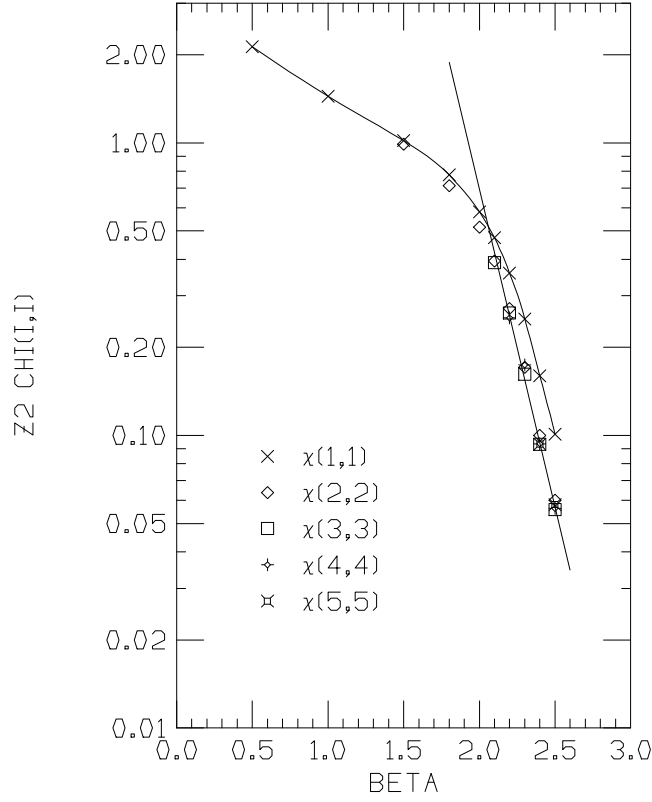,width=8.5cm,angle=90}}}
\caption[chi]{Creutz ratios from center-projected lattice
configurations.}
\label{chi}
\end{figure}

   The equality of Creutz ratios, starting at 2 lattice spacings,
means that the center projection sweeps away the short-distance,
$1/r$-type potential, and the remaining linear potential is revealed
already at short distances.  
This fact is quite apparent in Fig. 2, which displays
the data for $\chi(R,R)$ at $\b=2.4$ for the full theory (crosses),
the center projection (diamonds), and also for the $U(1)/Z_2$-projection
(squares).  The latter projection consists of the replacement $U \ra A/Z$ 
for the link variables.  We note that the center-projected data is virtually
flat, from $R=2$ to $R=5$, which means that the potential is linear in
this region, and appears to be the asymptote of the full theory.  It should
also be noted that abelian link variables with the center factored out, 
i.e. $U \ra A/Z$, appear to carry no string tension at all.

\begin{figure}
\centerline{\hbox{\psfig{figure=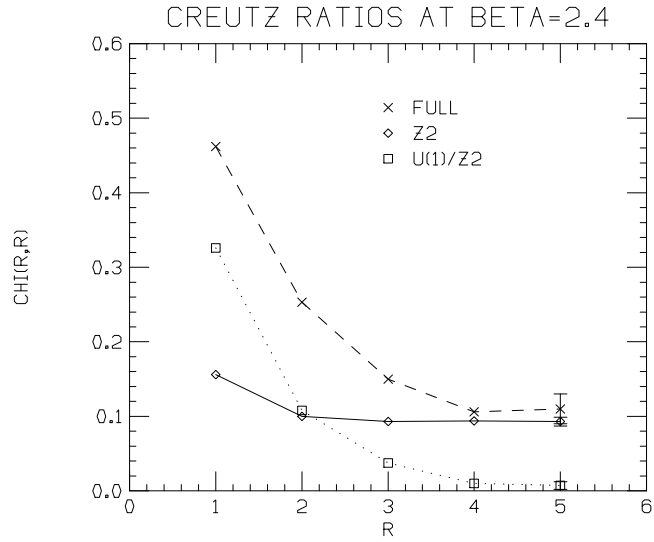,width=8.5cm,angle=90}}}
\caption[force]{Creutz ratios $\chi(R,R)$ vs. $R$ at $\b=2.4$, for
full, center-projected, and $U(1)/Z_2$-projected lattice 
configurations.}
\label{force}
\end{figure}

   Of course, one can also carry out finite temperature studies 
in the center projection.
Thus far, we have only computed Polyakov lines vs. $\beta$ on a
$6^3 \times 2$ lattice, and obtained the results shown in Fig. 3.
The deconfinement transition, signaled by a sudden jump in the value of
the Polyakov line, appears to occur at the value of $\beta$ appropriate
for $T=2$ lattice spacings in the time direction.

\begin{figure}
\centerline{\hbox{\psfig{figure=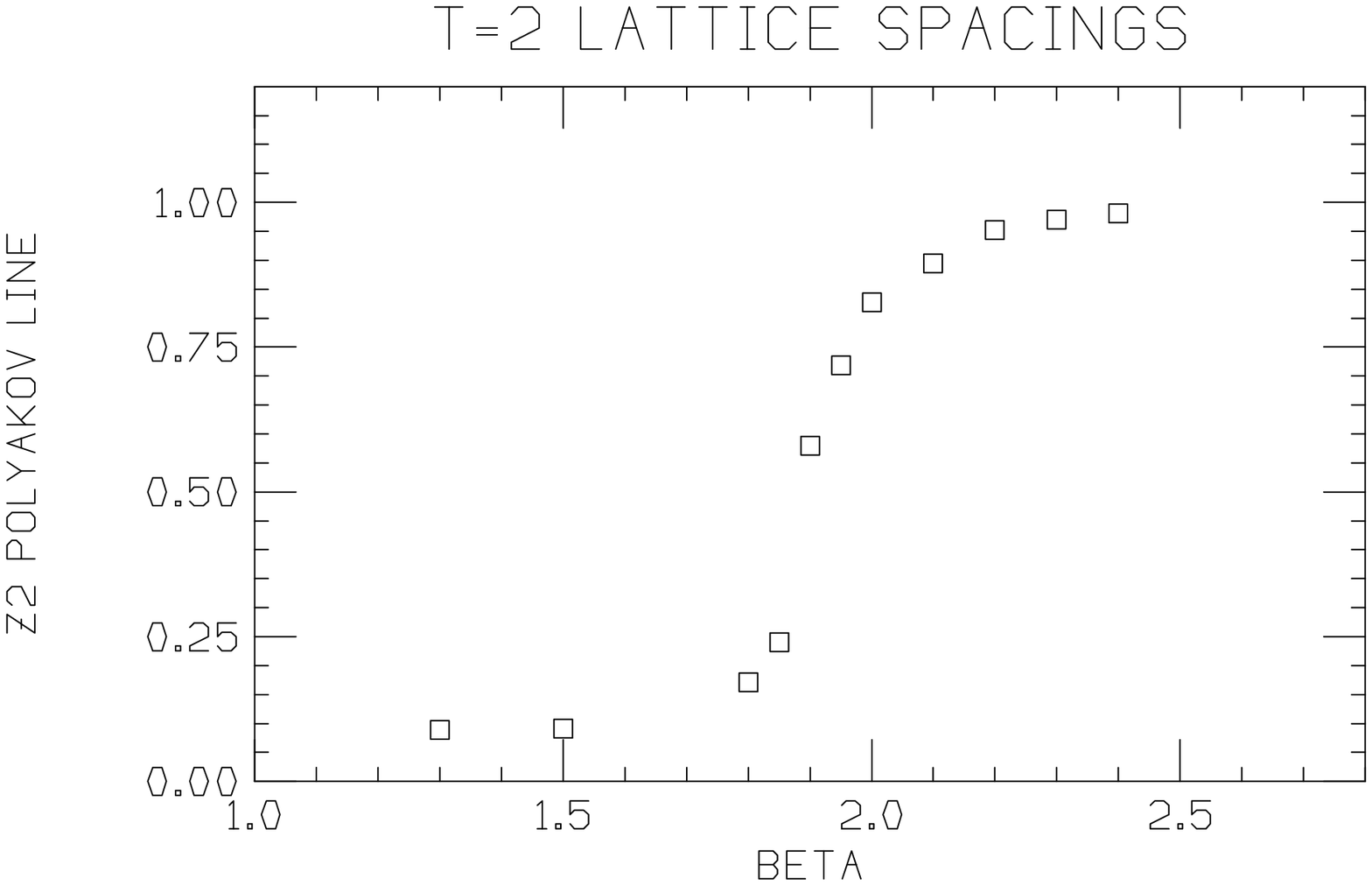,width=9.5cm,angle=90}}}
\caption[pol]{Polyakov lines vs. $\b$ in center projection, for
a $6^3 \times 2$ lattice.}
\label{pol}
\end{figure}

   It should be noted parenthetically that our definition of maximal center 
gauge is not the only possible definition.  A similar but not
identical gauge, leaving a remnant $Z_2$ symmetry, would be the gauge which 
maximizes
\beq
       \sum_x \sum_\m \{\mbox{Tr}U_\m(x)\}^2
\label{mZ2a}
\eeq
with center projection defined by
\beq
        Z = \mbox{sign}(\mbox{Tr}U)
\eeq
This version is more difficult to implement numerically, and has not yet been
studied.

   In any case, center dominance in maximal center gauge, as displayed in the 
figures above, does not necessarily imply that confinement is due to vortex
condensation.  In fact our initial view, expounded in 
ref. \cite{Us3}, was that since center dominance would appear to support 
a theory, namely vortex condensation, which is ``obviously wrong''
(for reasons discussed in section 4), 
its success only proves that neither center dominance nor abelian 
dominance are reliable indicators of the confinement mechanism.
The truncation of
degrees of freedom inherent in both the abelian and the center projections 
may easily do violence to the topology of the confining gauge fields. 
So the fact that the confining configurations of $U(1)$ gauge fields are 
monopoles, while
confining configurations in $Z_2$ gauge theory are condensed vortices,
does not necessarily imply that either type of configuration is 
especially relevant to the full, unprojected $SU(2)$ theory. 
 
    However, before stating with assurance that the $Z_2$ vortices
of the center-projected configurations have {\it nothing} to do
with confinement, there are certain checks that must be carried out.
It is here that we have encountered a surprise.

\section{The Detection of $Z_2$ Vortices}

    Consider a field configuration $U_\m(x)$, and any planar 
loop $C$.  As explained above, it is a simple matter to transform
to maximal center gauge, and then to examine each of the plaquettes
spanning the minimal area enclosed by loop $C$, in the corresponding
center projected configuration $Z_\m(x)$.  The number of plaquettes
computed with center-projected links, whose value is $-1$, corresponds
to the number of $Z_2$ vortex lines of the center-projected configuration
which pierce the minimal loop area.  We will refer to these $Z_2$ vortex 
lines of the 
center-projected configurations as ``projection-vortices,'' or just
``P-vortices,'' to distinguish them from the (hypothetical)
$Z_2$ vortices that might be present in the unprojected configurations.
As a Monte Carlo simulation proceeds, the number of P-vortices piercing
any given loop area will fluctuate.  The first question to ask is whether
the presence or absence of P-vortices in the projected configurations
is correlated in any way with the
confining properties of the corresponding unprojected configurations.

   To answer this question, we compute Creutz ratios $\chi_0(R,R)$ 
of Wilson loops $W_0(C)$, that are evaluated in a subensemble of Monte 
Carlo-generated configurations in which no P-vortex pierces the minimal 
area of loop $C$.  We stress that the full, unprojected link variables are 
used in computing the loop, and the center-projection is employed only
to select the data set.  In practice, having generated a lattice configuration
and fixed to maximal center gauge, one examines each rectangular loop of
a given size; those with no P-vortices piercing the loop are evaluated,
and those with a non-zero number are skipped.  Of course, by a trivial
generalization, we may compute Wilson loops $W_n(C)$, evaluated in
ensembles of configurations with any given number $n$ of P-vortices piercing 
the loop.    

   Figure 4 displays Creutz ratios $\chi_0(R,R)$ extracted from $W_0(C)$ loops,
as compared to the standard Creutz ratios with no such restriction, at 
$\beta=2.3$.  From this figure it is clear that, while the zero-vortex 
restriction makes little difference to the smallest 
loops, it makes a very big
difference to the Creutz ratios of the larger loops. It appears, in fact, 
that the aymptotic string tension of the zero-vortex loops vanishes 
altogether.\footnote{Error bars are much smaller for the no-vortex data
as compared to the full data; this is
why we can report meaningful results at larger $R$ for the no-vortex 
data than for the full data.}

\begin{figure}
\centerline{\hbox{\psfig{figure=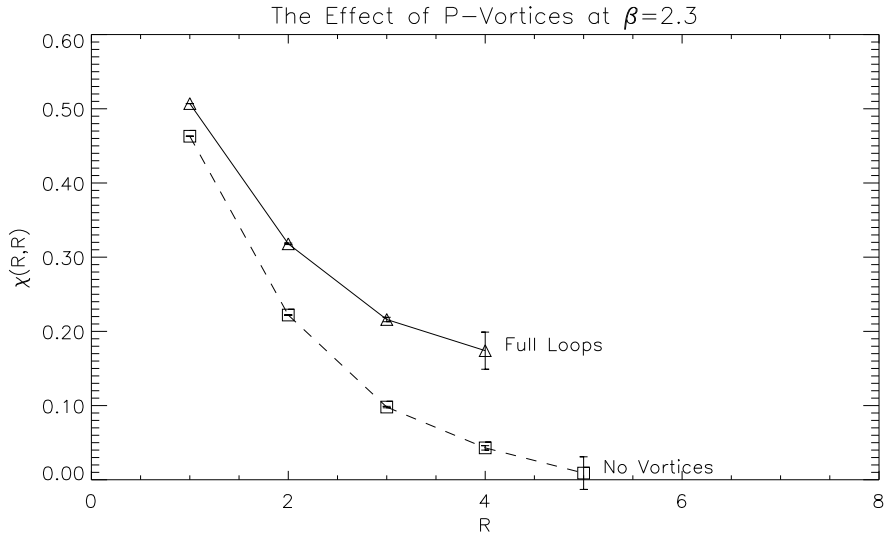,width=12.5cm,angle=0}}}
\caption[chi0]{Creutz ratios $\chi_0(R,R)$ extracted from
 loops with no P-vortices, as compared to the usual Creutz ratios
 $\chi(R,R)$, at $\b=2.3$.} 
\label{chi0}
\end{figure}
   If we presume (as most people do) that confinement is an effect 
associated with some particular type of field configuration - let us
call them the ``Confiners'' - then it would seem from Fig. 4 
that the presence or
absence of P-vortices in the center-projected configurations is strongly
correlated with the presence or absence of Confiners in the unprojected
configurations.  The next question is whether we can exclude the possibility
that these Confiners are actually $Z_2$ vortices.

   To address this question, assume for the moment that to each P-vortex
piercing a given loop, there corresponds a $Z_2$ vortex in the full,
unprojected field configuration piercing that loop.  This assumption has
the consequence that, in the limit of large loops,
\beq
       {W_n(C) \over W_0(C)} ~~ \longrightarrow ~~ (-1)^n
\label{pred_n}
\eeq

  The argument for eq. \rf{pred_n} goes as follows:  
In $SU(2)$ (as opposed to $Z_2$) lattice
gauge theory, the field strength of a vortex may be spread out in 
a cross-section, or ``core,'' of some finite diameter $D$ greater than 
one lattice spacing.   Outside
the core, the vector potential of each vortex can be represented by
a discontinuous gauge transformation. If a surface bounded
by loop $C$ is pierced by $n$ vortex lines, and if the cores of the
vortices do not intersect $C$, then the relevant gauge transformation,
at the point of discontinuity, has the property
\beq
       g(x(0)) = (-1)^n g(x(1))
\label{disc}
\eeq
where $x^\m(\tau),~\tau \in [0,1]$ parametrizes the closed loop $C$.
We can then decompose the vector potential
$A_\m^{(n)}(x)$ in the neighborhood of loop $C$ in terms of a
discontinuous gauge transformation $g(x)$, which represents the vortex
background near $C$, and a fluctuation $\d A^{(n)}_\m(x)$ such that
\beq
     A_\m^{(n)}(x) = g^{-1}\d A_\m^{(n)}(x) g + ig^{-1} \pa_\m g
\label{parts}
\eeq
The corresponding Wilson loop,
evaluated on the subensemble of configurations in which $n$ vortex
lines pierce loop $C$, would be
\bea
       W_n(C) &=& <\mbox{Tr}\exp[i\oint dx^\m A_\m^{(n)}]> 
\non \\
         &=&  (-1)^n <\mbox{Tr}\exp[i\oint dx^\m \d A_\m^{(n)}]>
\label{WW}
\eea
Of course, any vector potential in the neighborhood of loop
$C$ can be rewritten in the form \rf{parts}, so given some
criterion for identifying the number of $Z_2$ vortex lines piercing
loop $C$ (such as counting P-vortices), the question is whether 
this criterion, and the corresponding decomposition \rf{WW}, 
is physically meaningful.
A reasonable test is to see if the probability distribution of fluctuations
$\d A^n(x)$ is independent of the number of vortex lines piercing the
loop.  This test is based on the fact that,
in any local region of a large loop $C$, the effect of the
vortices is simply a gauge transformation.  Thus, providing the
fluctuations $\d A^n(x)$ have only short range correlations, their
distribution in the neighbourhood of loop $C$ should be unaffected by the 
presence or absence of vortex lines in the middle of the loop.
Therefore, if we have correctly isolated the vortex contribution,
\beq
<\mbox{Tr}\exp[i\oint dx^\m \d A_\m^{(n)}]> ~ \approx ~
      <\mbox{Tr}\exp[i\oint dx^\m \d A_\m^{(0)}> 
\eeq
for sufficiently large loops.  This immediately leads to eq. \rf{pred_n};
all that is needed is test this equation.  

    Figure 5 shows the ratio $W_1(C)/W_0(C)$ vs loop area,
for rectangular ($R\times R$ and $(R+1) \times R$) loops at $\beta=2.3$.
The simulations were performed on a 
$14^4$ lattice with 1000 thermalizing sweeps, followed by
8000 sweeps, with data taken every 10th sweep.  In order to give the 
loop $W_1(C)$ the greatest chance to lie outside the vortex core (assuming
it exists), $W_1(C)$ was evaluated in the subensemble of
configurations in which the single P-vortex is located in the center of
the loop.  The data seems perfectly consistent with eq. \rf{pred_n}, i.e. 
$W_1(C)/W_0(C) \ra -1$ as the loop area increases.

\begin{figure}
\centerline{\hbox{\psfig{figure=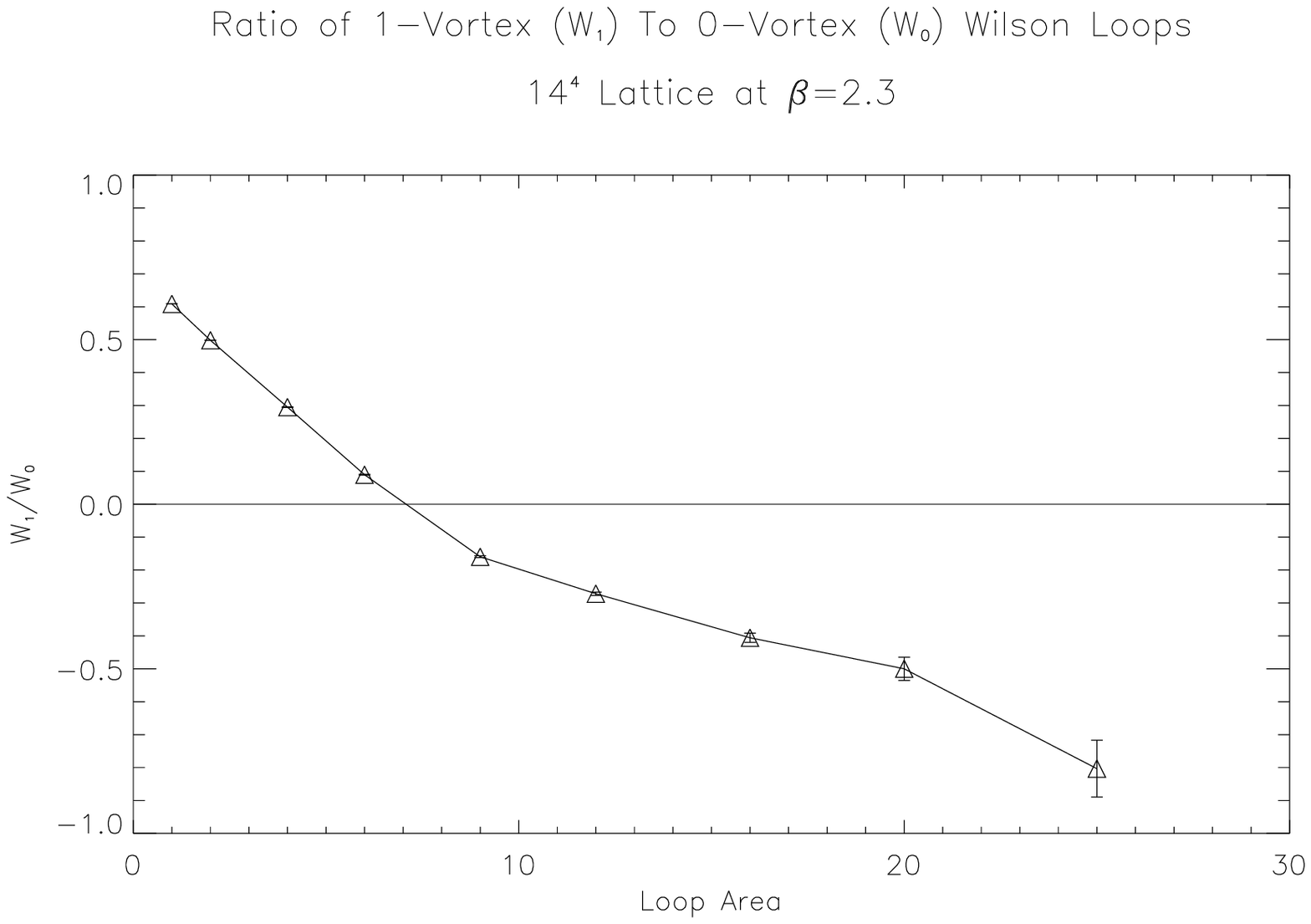,width=12.5cm,angle=0}}}
\caption[vtex1]{Ratio of the 1-Vortex to the 0-Vortex Wilson loops,
$W_1(C)/W_0(C)$, vs. loop area at $\b=2.3$.}
\label{vtex1}
\end{figure}
   
   Figure 6 shows the corresponding ratio $W_2(C)/W_0(C)$ vs loop
area.  In this case we have evaluated $W_2(C)$ in the subensemble of
configurations in which the two P-vortices lie inside a $2\times 2$
square in the middle of the loop.  As in the 1-vortex case, the idea is 
to keep the loop $C$ as far as a possible from the vortex cores, although
the cores themselves may overlap. Once again, 
the data seems to agree nicely with eq. \rf{pred_n}
for $n=2$, i.e. $W_2(C)/W_0(C) \ra +1$.

\begin{figure}
\centerline{\hbox{\psfig{figure=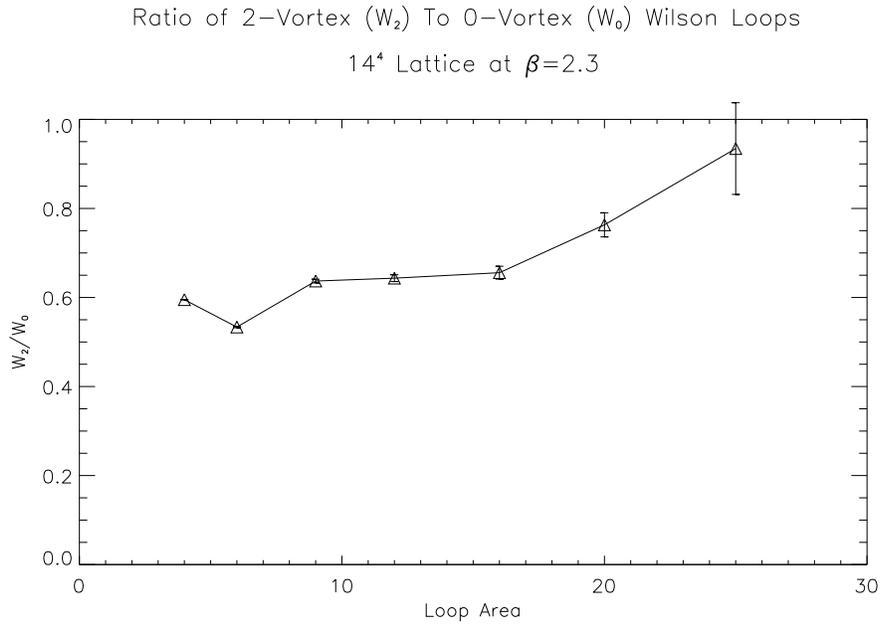,width=12.5cm,angle=0}}}
\caption[vtex2]{Ratio of the 2-Vortex to the 0-Vortex Wilson loops,
$W_2(C)/W_0(C)$, vs. loop area at $\b=2.3$.}
\label{vtex2}
\end{figure}

   Of course, the configurations that contain exactly zero (or 
exactly one, or two) P-vortices piercing a given loop become an 
ever smaller fraction of the total number of configurations, as
the loop area increases.  However, for increasingly large loops, one
would expect that the fraction of configurations with an even number
of P-vortices piercing the loop, and the fraction with an odd number
piercing the loop, approach one another.  This is indeed the case,
as can be seen from Figure 7.  

\begin{figure}
\centerline{\hbox{\psfig{figure=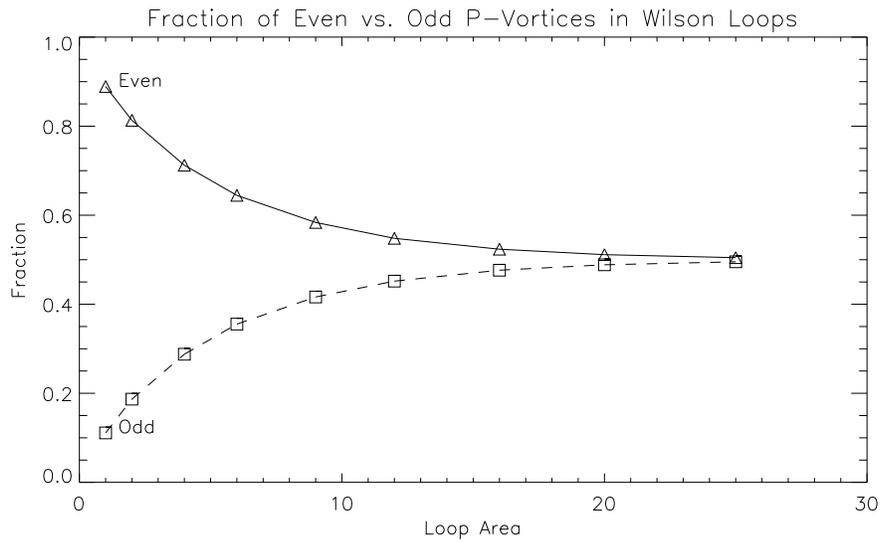,width=12.5cm,angle=0}}}
\caption[frac]{Fraction of link configurations containing 
even/odd numbers of P-vortices, at $\b=2.3$, piercing loops of various areas.}
\label{frac}
\end{figure}

    Let us define $W_{evn}(C)$ to be the Wilson loops evaluated in
configurations with an even (including zero) number of P-vortices
piercing the loop, and $W_{odd}(C)$ the corresponding quantity for
odd numbers.\footnote{In evaluating $W_{evn}$ and $W_{odd}$, we 
make no special restriction on the location of the P-vortices within
the loop.}  
According to eq. \rf{pred_n}, $W_{evn}(C)$ and
$W_{odd}(C)$ should be of opposite sign, for large loop area.
Moreover, according to the vortex condensation picture, the area
law for the full loop $W(C)$ is due to fluctuations in the 
$\pm 1$ factor, coming from fluctations in even/odd numbers of
vortices piercing the loop.  If this is the case, then neither
$W_{evn}(C)$ alone, nor $W_{odd}(C)$ alone, would have an area law,
but only the weighted sum
\beq
      W(C) = P_{evn}(C) W_{evn}(C) + P_{odd}(C) W_{odd}(C)
\label{wsum}
\eeq
where $P_{evn}$ and $P_{odd}$ are the fractions of configurations, 
shown in Fig. 7, with even/odd numbers of P-vortices piercing the loop. 
For large loops, $P_{evn} \approx P_{odd} \approx 0.5$.

\begin{figure}
\centerline{\hbox{\psfig{figure=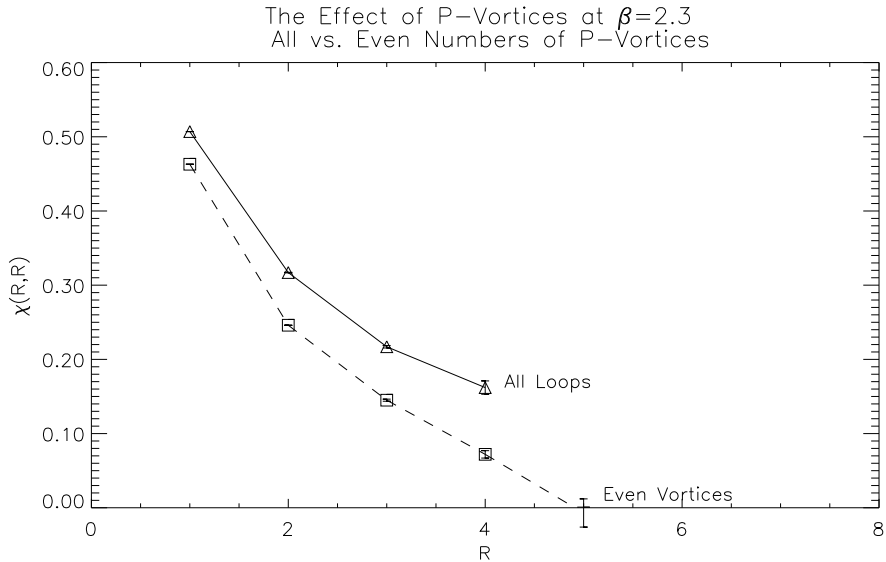,width=12.5cm,angle=0}}}
\caption[chieo]{Creutz ratios $\chi_{ev}(R,R)$ extracted from Wilson
 loops $W_{evn}(C)$, taken from configurations with even numbers of 
 P-vortices piercing the loop.  The standard Creutz ratios
 $\chi(R,R)$ at this coupling ($\b=2.3$) are also shown.}
\label{chieo}
\end{figure}

   Figure 8 shows the Creutz ratios extracted from $W_{evn}(C)$, 
compared to the standard Creutz ratios at $\b=2.3$.  The figure is
qualitatively quite similar to Fig. 4, but here it should be emphasized
that the data set used to evaluate $W_{evn}(C)$ is not a small minority of
configurations (as it is for $W_0(C)$ for large loops), but constitutes at
least half the configurations.  The asympotic string tension,
extracted from these configurations, appears to vanish.  

   Figure 9 shows the values of $W_{evn}(C),~W_{odd}(C),~W(C)$ vs. loop
area, for the larger loops.  As expected from eq. \rf{pred_n}, $W_{evn}$
and $W_{odd}$ have opposite signs.  The full Wilson loop $W(C)$ has
a positive sign, but is substantially smaller, at loop area $\ge$ 20,
than either of its two components.  If this behavior persists
at still greater areas, then the area law falloff of
a Wilson loop $W(C)$ is due to a very delicate cancellation between much
larger positive and negative components, associated with even and
odd numbers of P-vortices respectively.  Neither $W_{evn}(C)$ nor
$W_{odd}(C)$, by itself, would appear to have an area law.

\begin{figure}
\centerline{\hbox{\psfig{figure=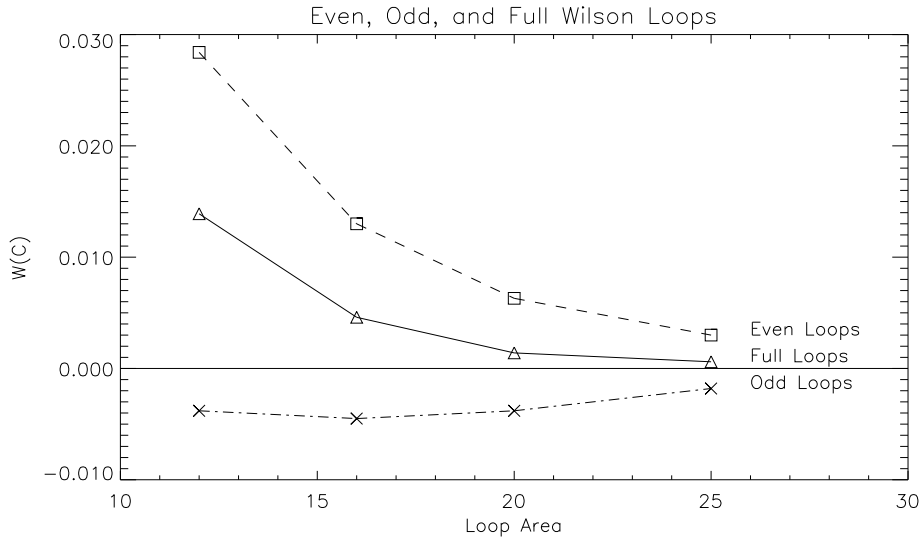,width=12.5cm,angle=0}}}
\caption[vtexeo]{ Wilson loops $W_{evn}(C), ~W_{odd}(C)$ and $W(C)$
at larger loop areas, taken from configurations with even numbers
of P-vortices, odd numbers of P-vortices, and any number of P-vortices,
respectively piercing the loop.  Again $\b=2.3$.}
\label{vtexeo}
\end{figure}

\section{Against Vortices}

    The data presented in the previous section suggests that $Z_N$
vortices play a crucial role in the confinement process, and that
condensation of such vortices (as proposed in refs. \cite{Poul,tH1,Mack})
may be the long-sought confinement mechanism.  On the other hand, there
are some serious objections which can be raised against this mechanism.
We have raised these objections repeatedly, in connection with the
abelian-projection theory \cite{Us1,Us2,Us3}, and they apply with even more 
force to the vortex-condensation theory.  

    The difficulties are all associated with 
Wilson loops in higher group representations.  First of all, there
is a problem  concerning the large-N limit \cite{Me}.  A Wilson loop for
quarks in the adjoint representation of an $SU(N)$ gauge group is 
unaffected by the discontinuous gauge transformations associated with $Z_N$  
vortices; it follows that fluctuations in the number of 
such vortices cannot produce an area law
for adjoint loops.  On the other hand, it is a consequence of factorization
in the large-N limit that, at $N=\infty$, the string tension of the
adjoint loop $\s_{Adj}$ is simply related to the string tension of the
$\s_{fund}$ of the fundamental loops
\beq
         \s_{Adj} = 2 \s_{fund}
\eeq
In addition, the existence of an adjoint string tension does not appear
to be just a peculiarity of the large-N limit.  It has been found in
numerous Monte Carlo investigations, for both the $SU(2)$ and $SU(3)$
gauge groups in both three and four dimensions, that there is an 
intermediate distance regime, from the onset of confinement to the
onset of color screening, where 
\beq
        {\s_r \over \s_{fund} } \approx {C_r \over C_{fund}}
\eeq
where $C_r$ is the quadratic Casimir of representation $r$ 
\cite{AOP,Casimirs,Manfried}.
Again, it is hard to see how vortex condensation would account for
this ``Casimir scaling'' of string tensions in the intermediate distance
regime.

   From these considerations, it is clear that the
``Confiners,'' whatever they may be, must produce rather different effects
in different distance regimes.  In $SU(2)$ gauge
theory, in the intermediate distance regime,
the Confiners should supply string tensions compatible with
\beq
      {\s_j \over \s_\oh} \approx {4 \over 3} j(j+1)
\eeq
while, from the onset of
color-screening and beyond, they should produce asymptotic values
\beq
       \s_j = \left\{ \begin{array}{cl}
                       \s_\oh & j = ~~\mbox{half-integer} \\
                         0    & j = ~~\mbox{integer} \end{array}
              \right.
\eeq
Both the vortex-condensation and abelian-projection theories are
compatible with the latter condition on asymptotic string tensions,
but do not explain Casimir scaling at intermediate distances.

   It is entirely possible that $Z_N$ vortices (or, for that matter,
magnetic monopole configurations) have something to do with the
confinement mechanism at distance scales beyond the onset
of color-screening.  But it is also possible that the data of the
previous section could be misleading in some way, and it is worth considering
how that could happen.  Let us rewrite eq. \rf{wsum} in the form
\beq
       W(C) = \D P(C) W_{evn}(C)  + P_{odd}(C) \D W(C)
\label{exact}
\eeq
where
\bea
       \D P(C) &\equiv& P_{evn}(C) - P_{odd}(C)
\non \\
       \D W(C) &\equiv& W_{evn}(C) + W_{odd}(C)
\eea
In the center projection, $W_{evn}=1$ and $W_{odd}=-1$ so that
\beq
      W_{cp}(C) = \D P(C)
\eeq
If $Z_2$ vortices are the confiners, then, as in the center projection,
the area law is due to random fluctuations in the number of vortices
piercing the loop. It would then be the term proportional
to $\D P(C)$ in eq. \rf{exact} which accounts for the asymptotic
string tension in the full theory.
Asymptotically, $P_{evn} \approx P_{odd} \approx \oh$, so that   
\beq
       W(C) \ra \D P(C) W_{evn} + \oh \D W(C)
\label{W}
\eeq
To really establish that $Z_2$ vortices are the origin of the asymptotic
string tension, we need (among other things) to establish that
\beq
       \D P(C) \sim \exp[-\s_{cp} A(C)]
\eeq
with
\beq
             \s_{cp} = \s_{fund}
\eeq
where $\s_{cp}$ is the string tension of the fundamental representation
in center projection.  Proper scaling of $\s_{cp}$ with respect to
$\b$ is a necessary but not sufficient condition for this equality, and
this is one way that the previous data might be misleading.
If it should turn out that
\beq
            \s_{cp} > \s_{fund}
\eeq
then the first term on the rhs of eq. \rf{W} would become 
negligible, asymptotically, compared to $W(C)$, and the asymptotic 
string tension would have to be due to $\D W(C)$.  

  From Fig. 2, and the scaling apparent in Fig. 1, it would appear that
$\s_{fund}$ and $\s_{cp}$ are not very different.  In this connection, it is
instructive to compare Fig. 2 with an analogous calculation in compact
$QED_3$.  In $QED_3$ it is also possible to define a ``maximal $Z_2$
gauge,'' via eq. \rf{maxZ2}, and a ``$Z_2$ projection'' 
(the term ``center projection''
would be a misnomer here) according to eq. \rf{cp}.  A sample
result for lattice $QED_3$ is shown below in Fig. 10.  
This simulation was run on a $30^3$
lattice at $\beta=2.2$, with $5000$ thermalizations followed by $10000$
sweeps, with data taken every $10$th sweep.  In this case, Creutz ratios
$\chi_{cp}(R,R)$ for the projected data appear to be approximately 
$40\%$ larger than
the Creutz ratios $\chi(R,R)$ for the unprojected data, and 
the two quantities don't appear to be converging for larger loops.  

   The agreement between projected and unprojected Creutz ratios appears 
to be substantially better in the $D=4$ $SU(2)$ theory
than in compact $QED_3$, despite the fact that
$SU(2)$ is a larger group than $U(1)$.
On the other hand, the data presented in section 3 for the non-abelian
theory was obtained on workstations, not supercomputers.
A state-of-the art string tension calculation
aimed at a better quantitative comparison of $\s$ and
$\s_{cp}$,  and perhaps also a study of alternate versions of the
maximal center gauge (such as \rf{mZ2a}), is certainly called for.

\begin{figure}
\centerline{\hbox{\psfig{figure=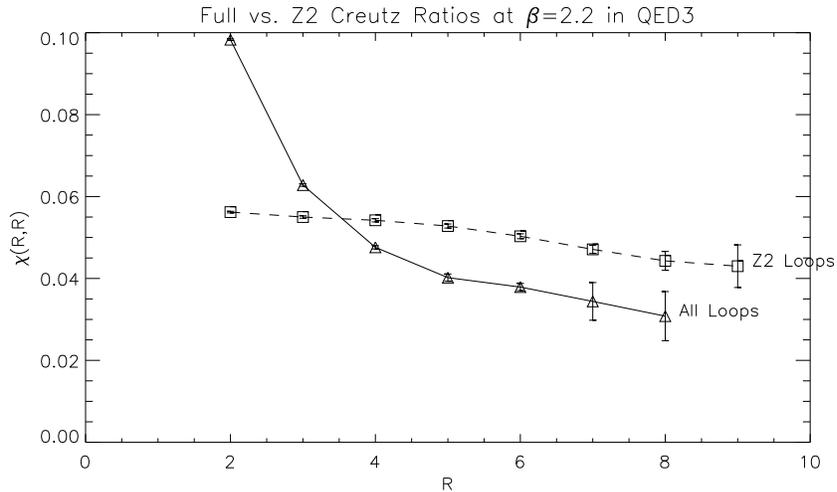,width=11.5cm,angle=0}}}
\caption[u1cp22]{Test of $Z_2$ dominance in compact $QED_3$: $Z_2$ Projected 
vs. Full Creutz ratios at $\b=2.2$ on a $30^3$ lattice.}
\label{u1cp22}
\end{figure}

\section{Conclusions}

   None of the evidence gathered so far is conclusive, 
although it does seem to point in a certain direction.  
We have found that:

\begin{description}

\item{1.} Center-projected link variables have the property of
``center dominance'' in maximal center gauge.  In this gauge it is the
sign alone, of the real part of the abelian-projected link, which appears 
to carry most of the information about the asymptotic string tension.

\item{2.} Vortices in the center-projected configurations (``P-vortices'')
appear to be strongly correlated with the presence or 
absence of confining field
configurations in the full, unprojected field configurations.
When Wilson loops are evaluated in an ensemble of configurations
which do not contain P-vortices within the loop, the asymptotic string
tension disappears.

\item{3.} Wilson loops $W_0(C),~W_1(C),~W_2(C)$, evaluated in ensembles
of configurations containing respectively zero, one, or two P-vortices 
inside the loop, in the corresponding center projection, behave as
though they contained zero, one or two $Z_2$ vortices in the full,
unprojected configuration.  That is, $W_1/W_0 \ra -1$, and $W_2/W_0 \ra +1$,
as the loop area increases.

\end{description}

   If the Yang-Mills vacuum is dominated by $Z_2$ vortices, as this
data would seem to suggest, it raises many puzzling questions.  Foremost 
among these is how to account for the existence and Casimir scaling of 
the adjoint string tension.  Because of the existence of the adjoint tension, 
we think it unlikely that fluctuations in the number and location of 
$Z_2$ vortices can give a complete account of 
the confinement mechanism in the intermediate distance
regime. Such vortex fluctuations {\it could} be decisive asymptotically; 
further work will be needed to find out.

   Perhaps the most urgent need is to repeat all of the calculations 
reported here for the case of an $SU(3)$ gauge group.  If there is
center dominance (with $\s \approx \s_{cp})$, and if the presence of 
P-vortices is correlated with the 
magnitude of the string tension, and especially if
\beq
       {W_1(C) \over W_0(C)} ~~ \longrightarrow ~~ e^{2\pi i/3}
\eeq
for corresponding P-fluxons with one unit $e^{2\pi i/3}$ of center
flux, then we believe that the combined evidence in favor of some version
of the $Z_N$ vortex condensation theory would become rather compelling.

\vspace{33pt}

\noindent {\Large \bf Acknowledgements}

\bigskip

  J.G. is grateful for the hospitality of the Niels Bohr
Institute, where some of this work was carried out. He would
also like to thank J. Ambj{\o}rn and T. H. Hansson for discussions.
This research was supported in part by an EC HMC Institutional Fellowship
under Contract No. ERBCHBGCT930470 (L.DD.), the 
Hoch\-schu\-l\-ju\-bi\-l\"au\-mssti\-f\-tung
der Stadt Wien H-00114/95 (M.F.), the
U.S. Dept. of Energy, under Grant No. DE-FG03-92ER40711 (J.G.),
and the Slovak Grant Agency for Science, Grant No. 2/1157/94
(\v{S}.O.).  Support was also provided by the Director,
Office of Energy Research, Office of Basic Energy Services, of the
U.S. Department of Energy under Contract DE-AC03-76SF00098.


\newpage

\begin{thebibliography}{xx}
\bibitem{tH2} G. 't Hooft, Nucl. Phys. B190 [FS3] (1981) 455.
\bibitem{Suzuki} T. Suzuki and I. Yotsuyanagi, Phys. Rev. D42 (1990) 4257; \\
 S. Hioki et al., Phys. Lett. B272 (1991) 326.
\bibitem{KLSW} A. Kronfeld, M. Laursen, G. Schierholz, and
U.-J. Wiese, Phys. Lett. B198 (1987) 516.
\bibitem{Poli} M. Polikarpov, to appear in the
Proceedings of LATTICE 96, archive: hep-lat/9609020.
\bibitem{Poul} H. B. Nielsen and P. Olesen, Nucl. Phys. B160 (1979) 380.
\bibitem{tH1} G. 't Hooft, Nucl. Phys. B153 (1979) 141.
\bibitem{Mack} G. Mack, in {\sl Recent Developments in Gauge Theories},
edited by G. 't Hooft et al. (Plenum, New York, 1980).
\bibitem{Yoneya} T. Yoneya, Nucl. Phys. B144 (1978) 195.
\bibitem{Tomboulis} T. Kov\'{a}cs and E. Tomboulis, to appear in the
Proceedings of LATTICE 96, archive: hep-lat/9607068.
\bibitem{AO}  J. Ambj{\o}rn and P. Olesen, Nucl. Phys. B170 [FS1]
(1980) 265.
\bibitem{Us3} L. Del Debbio, M. Faber, J. Greensite, and 
{\v S}. Olejn\'{\i}k, to appear in the
Proceedings of LATTICE 96, archive: hep-lat/9607053.
\bibitem{Us1} L. Del Debbio, M. Faber, and J. Greensite, Nucl. Phys. B414
(1994) 594.
\bibitem{Us2} L. Del Debbio, M. Faber, J. Greensite, and 
{\v S}. Olejn\'{\i}k, Phys. Rev. D53 (96) 5891.
\bibitem{Me} J. Greensite and M. B. Halpern, Phys. Rev. D27 (1983) 2545.
\bibitem{AOP} J. Ambj{\o}rn, P. Olesen, and C. Peterson, Nucl. Phys. B240
[FS12] (1984) 189; 533.
\bibitem{Casimirs} G. Poulis and H. Trottier, archive: hep-lat/9504015; \\
C. Michael, Nucl. Phys. B (Proc. Suppl.) 26 (1992) 417;
Nucl. Phys. B259 (1985) 58.
\bibitem{Manfried}
N. Cambell, I. Jorysz, and C. Michael, Phys. Lett. B167 (1986) 91; \\
M. Faber and H. Markum, Nucl. Phys. B (Proc. Suppl.) 4 (1988) 204; \\
M. M\"uller, W. Beirl, M. Faber, and H. Markum,  Nucl. Phys. B
(Proc. Suppl.) 26 (1992) 423.
\end{thebibliography}
\end{document}